# Anomalous Hysteresis Behavior in Sputter-deposited Ultrathin Films of Amorphous- CoFeB Alloy

Baisali Ghadai, Kirti Kirti, Abinash Mishra, Sucheta Mondal*

Physics Department, Shiv Nadar Institution of Eminence, Greater Noida, Delhi NCR,

UP-201314, India

*sucheta.mondal@snu.edu.in

## Abstract

Thin amorphous-$Co_{20}Fe_{60}B_{20}$ (a-CFB) is deposited by rf-magnetron sputtering on a self-oxidized Si (100) substrate with different film thicknesses ranging from 0.7 nm to 20 nm. The 5-nm-thick a-CFB film is capped with a W layer for comparison. The surface morphology is investigated by using the atomic force microscopy technique. The low roughness of all the films' surface indicates uniformity, moderate corrosion resistance, and good structural quality. The X-ray diffraction spectra reveal the amorphous nature of the CFB layer, while the W capping is of mixed $\alpha - \beta$ phase in the experimental thickness regime. In-plane and out-of-plane hysteresis loops obtained from the vibrating sample magnetometry technique show a transition from an 'upright S' to 'nearly rectangular' shape *via* a 'completely inverted' profile. A self-sustained tilted magnetic anisotropy is stabilized in a seed-free environment based on the direct substrate-to-magnet interaction. The interface anisotropy is estimated to be 0.06 erg/cm$^2$. The complex anisotropic behavior originates from the interplay between interface anisotropy, conventional shape anisotropy, growth-induced anisotropies, and inhomogeneity-induced anisotropies. In essence, effective anisotropy is responsible for the anomalous hysteresis behavior observed in these films, and this work might provide valuable insights to improve the functionalities of amorphous soft magnetic alloys.

## 1. Introduction

Amorphous CoFeB (a-CFB hereafter) thin films deposited by physical vapor deposition processes remained underrated for years after this was reported for the first time. In 1978, Heiman et al. reported growth-induced improvement in magnetic properties in a-$Co_{74}F_6B_{20}$ thin films deposited on quartz substrates by using the rf diode sputtering technique [1]. These films possessed high permeability, high saturation magnetization ($M_s$), very low coercivity, negligible magnetostriction, moderate corrosion resistance, etc. Similar observations were reported by A. Brunsch in 1979 [2]. Magnetic anisotropy of such soft magnetic films prepared by the field annealing method (≈1 $\mu$m thickness) was estimated from the difference between the magnetization curves with magnetic field swept along the easy and hard axes [3]. Extensive research on magneto-resistive (MR) devices emerged by that time. Such devices require a high MR ratio, low anisotropy, low magnetostriction, high stability, etc. All these qualities were not observed in a-CFB thin films (as those might differ based on the growth conditions), as reported by U Dibbern while proposing MR-based magnetic field sensors [4]. Researchers found thermal and rotating field annealing useful in improving some of these qualities avoiding growth-induced anisotropy-related complexities in sputter-deposited a-CFB films [5]. A relatively high MR was reported for a 4 $\mu$m thick a-CFB single layer in 1995 [6]. Later, the importance of a-CFB thin layers embedded into spin-valve devices (that could be integrated into read heads) was highlighted. A high GMR ratio and higher field sensitivity were obtained for Co/Cu/a-CFB stack by improving the magnetic softness of the free layer compared to polycrystalline NiFe layers used previously [7]. 1996 onwards sputter-deposited ultrathin a-CFB layers became reliable candidates for integration into spin-valve multilayers as the fixed and free layers for sensor applications [8–10]. At the beginning of this century, there was a surge in research with a-CFB-based magnetic tunnel junctions (MTJs) and their possible integration in magnetoresistive random-access memory (MRAM) for nonvolatile data storage applications [11]. Researchers achieved a very high TMR ratio (upto 130% at room temperature) and a high spin-polarization for the free a-CFB layer by tuning the deposition conditions, particularly by applying in-situ annealing. These MTJs can operate with low charge current densities and have a low $M_sV$ (V is the volume) product suitable for ultra-dense MRAM architecture. It is pertinent to mention that the memory state in MRAM is maintained by the relative direction of magnetization between the layers, and not by the charge current supplied to it, unlike oxide-based flash memory. MRAM has three major kinds: Toggle MRAM, spin transfer torque (STT) MRAM, and spin orbit torque (SOT) MRAM. The most relevant reports highlighting the involvement of ultrathin a-CFB layers in those architectures are discussed chronologically here. In 2005, Kim et al. reported the first toggle MRAM by using CFB as the primary element in the MTJ stacking without emphasizing the role of amorphous nature in controlling the device functionality [12]. Later, in 2006, a-CFB based MTJ was reported. With an increase in Boron (B) concentration of the free layer, an increase in MR, subsequently, a further decrease was observed for the toggle MRAM stack. This is attributed to the crystalline-to-amorphous phase transition resulting in the highest MR for the stack at the transition point, which can be further tuned by in-situ annealing [13]. However, a dedicated amorphous phase was found to be useful neither for further improvement in the operation of toggle MRAM nor for other kinds of MRAM (i.e., STT-MRAM and SOT MRAM), as the stability of perpendicular magnetic anisotropy (PMA) was an issue [14,15]. Recently, there has been renewed interest in a-CFB films. Achievement of ultralow switching current is reported in a-CFB-based all-amorphous spintronic devices [16]. Thin a-CFB layer deposited on TI films has been reported as an excellent THz emitter [17,18] candidate due to robustness to oxidation [19] and high spin-polarization [20].

Due to the huge application potential of a-CFB films, the structural and magnetic properties have been extensively studied over the years. It is important to mention that this alloy exhibits in-plane (IP) magnetization behavior with the hard axis lying towards the out-of-plane (OOP) direction when the film thickness is few tens of nm [21]. This behavior is dominated by conventional shape anisotropy when crystalline anisotropy is absent, and other induced anisotropies are dormant. For the ultrathin films with thickness in the sub-nm range, the interfacial anisotropy effect surpasses the shape anisotropy (SA) and stabilizes the magnetization in the OOP direction [22]. In the ultrathin to thin regime, a-CFB films may exhibit various other anisotropies (e.g., interface-induced, growth-induced, heat-induced, field-induced, stress-induced, etc.) [23–27]. The interplay between these anisotropies gives rise to an effective IP uniaxial magnetic anisotropy (UMA). In some cases, it results in unusual magnetization reversal phenomena. A volume UMA in a-CFB (thickness range: 3.5-20 nm, film composition and applied elastic strain are varied as well) is seeded due to growth conditions during film deposition, which has a microstructural origin like the previously proposed "bond-orientational" anisotropy (BOA) model for rare-earth magnets. Due to competition between UMA and the SA, hysteresis loops with two-stage magnetization reversal are obtained [24]. Similar hysteresis behavior was reported by R. Lavrijsen, however in CFB layer seeded with Ta/Pt [28].

Despite the humongous effort made by the scientific community to understand the magnetic properties of a-CFB thin films, we have identified gaps in the literature that require experimental validation. No reports compile the systematic deconvolution of all the possible anisotropy components governing the magnetization reversal behavior in a-CFB thin films for sub-nm to nm thickness range. The quantification of thickness-dependent effective anisotropy by comparing the hysteresis loops in the IP and OOP configurations is also missing for such systems. Importantly, stabilization of effective magnetization with substantial tilt in the OOP direction for sub-nm a-CFB films grown solely on self-oxidized Si substrate without a strategic seeding layer has not been attempted so far. Moreover, apart from the two-stage magnetization reversal discussed above, no reports are available on negative remanence observed in a completely 'inverted' hysteresis loop for such films, which is of fundamental importance.

In this study, we have deposited a-CFB films in ultrathin-to-thin regime with thicknesses varying from 0.7-20 nm on Si (100) substrate. The uncapped Si/CFB ($t_{CFB}$ = 5 nm) are also compared with capped Si/CFB ($t_{CFB}$ = 5 nm)/W ($t_W$ = 1-10 nm). The IP and OOP hysteresis loops obtained from vibrating sample magnetometry (VSM), are used to understand the magnetization reversal at various thicknesses. The individual anisotropy contributors responsible for the magnetization behavior at different thicknesses are identified. The interface and bulk anisotropy contributions to the effective anisotropy are calculated to be 0.06 erg/cm$^2$ and 9 × 10$^6$ erg/cc. Also, the phenomena responsible for tilted magnetization and characteristic two-stage magnetization reversal resulting in so called "inverted loops" with negative remanence are deciphered. This work provides critical insights towards enhancing the material properties of CFB like alloys, which can act as building block for the efficient next generation spintronic devices.

## 2. Experimental details

Ultrathin-to-thin films of $C_{20}Fe_{60}B_{20}$ (CFB) with thickness $t_{CFB}$ = 0.7, 1, 3, 5, 7, 10, 20 nm and Tungsten (W) with thickness: $t_W$ = 1, 3, 5, 7, 10 nm are deposited at room temperature on p-type Si (100) substrate using high-vacuum radio frequency (rf) (13.56 MHz) magnetron sputtering system designed by Excel Instruments. The base pressure of the chamber is maintained below 7 ×

$10^{-7}$ Torr. CFB is deposited at an Argon working pressure of 5 mTorr using a rf power of 90 Watt, while W is deposited at 7 mTorr with power of 80 Watts. Argon flow rate is 40 SCCM for both. To ensure uniformity, all the deposition conditions are carefully optimized and kept identical across all the samples of interest. Based on these optimized conditions, CFB ($t_{CFB}$ = 5 nm) with W capping layer of varying thickness ($t_W$ = 1, 3, 5, 7, 10 nm) is deposited, resulting in Si/CFB ($t_{CFB}$ = 5 nm)/W ($t_W$) stack configurations. The deposition flux approaches the substrate at an angle of ≈$45^0$ relative to the substrate normal, i.e. referred to as the incidence angle, as shown in Fig. 1(a), while the substrate is rotated at 2 rpm. The uncapped films (see Fig. 1(b)) are used for structural, electrical, and magnetic characterizations. The contribution from the top oxide layer is negligible in altering the sample properties demonstrated in this paper. Each sample has lateral dimensions of 1 × 1 cm². The thickness of the samples is determined by using Profilometer, allowing for the optimization of thickness, indicating a deposition rate of approximately 0.6 Å/sec, and 4 Å/sec for CFB and W, respectively (Fig. 1S). To know about the surface topography, AFM measurements were conducted in intermittent contact (tapping) mode using Oxford Instruments MFP-3D ORIGIN Asylum research by taking a scan over 5µm × 5 µm area and analyzed using Gwyddion software. The crystallinity of Si/CFB ($t_{CFB}$), Si/W ($t_W$), and Si/CFB ($t_{CFB}$ = 5 nm)/W ($t_W$) thin films is investigated by Grazing-incidence X-ray diffraction (GI-XRD) measurements using Cu-Kα (0.154187 nm) radiation on a Rigaku Smart Lab SE diffractometer operated at 3kWatt. All measurements are performed at room temperature within the range 2θ = 30° to 80° at a scan rate of 10°/min with 0.02° resolution. The background-subtracted XRD spectra for Si/CFB ($t_{CFB}$), Si/W ($t_W$), and Si/CFB ($t_{CFB}$ = 5 nm)/W ($t_W$) thin films are presented in Fig. 2. Peak intensities with respect to the baseline are retained as obtained during the XRD measurement in these plots.

Static magnetic properties are studied by using VSM attached to a Quantum Design Dynacool PPMS instrument. The external magnetic field is swept between 2T and -2T with 20 Oe field resolution near the reversal. The measurements are performed at room temperature for all the samples. The hysteresis loops are obtained in both parallel and perpendicular directions with the magnetic field.

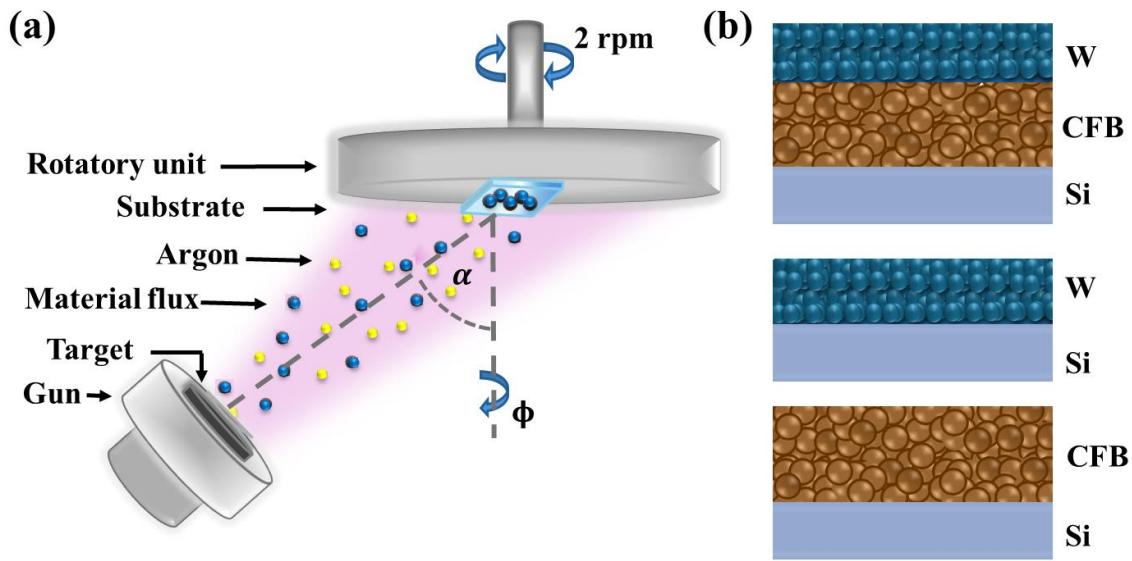

Figure 1: (a) Schematic view of the deposition geometry, where $\alpha$ and $\varphi$ are angles of incoming flux with the substrate normal and substrate rotation respectively. (b) The sample stacks for the Si/CFB, Si/W and Si/CFB/W are presented. The random and ordered arrangement of atoms are depicted for amorphous and crystalline materials.

## 3. Results and discussion

### 3.1 Structural and topological properties

Figure 2 illustrates the AFM images for Si/CFB ($t_{CFB}$) and Si/CFB ($t_{CFB}$ = 5 nm)/W ($t_W$) heterostructures with different thicknesses. It shows that the film surfaces are uniform. The average topological roughness is obtained from these images as listed below in Table 1. It ranges between 0.09 and 0.24 for Si/CFB ($t_{CFB}$). For Si/CFB ($t_{CFB}$ = 5 nm)/W ($t_W$), this value is between 0.13 and 0.3, which is slightly higher than Si/CFB ($t_{CFB}$). The variation in surface roughness is found to be significantly small when measured at different regions of the same sample, irrespective of the capping layer as well. The surface topography of other samples is presented on the supplementary material (Fig. 2S). The AFM image for the Si substrate is shown in Fig. 3S.

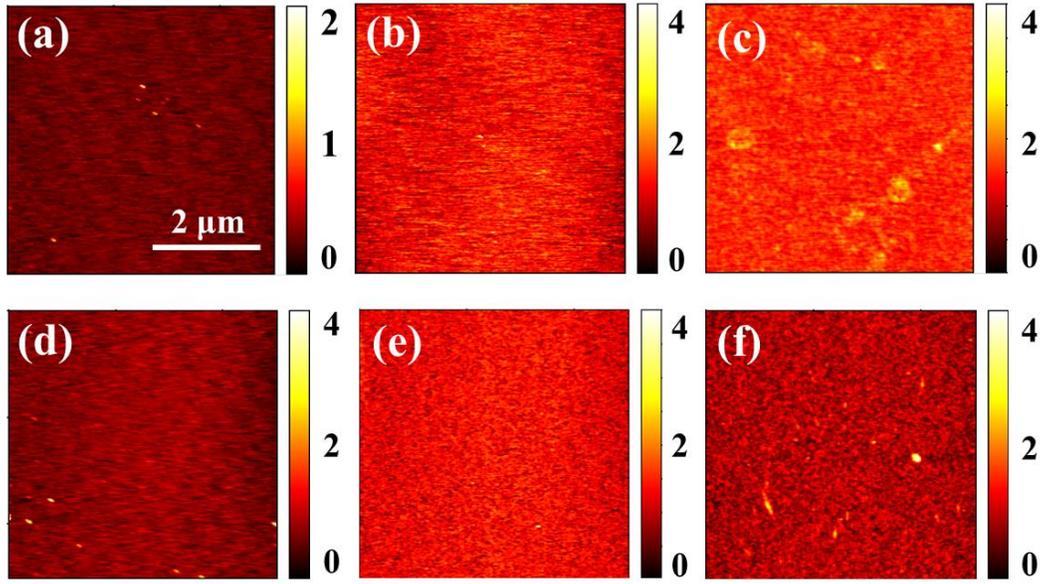

Figure 2: AFM images of (a-c) Si/CFB ($t_{CFB}$ = 1, 5 and 10 nm), and (d-f) Si/CFB ($t_{CFB}$ = 5 nm)/W ($t_W$ = 1, 7 and 10 nm). The scan area shown is 5μm × 5 μm for all the samples. The length scale is given for reference. Height profiles are depicted by colour bars.

**Table 1:** Average roughness variation for Si/CFB ($t_{CFB}$) and Si/CFB ($t_{CFB}$ = 5 nm)/W ($t_W$) films.

| Sample | Average roughness (nm) |
| --- | --- |
| CFB 1 | 0.09 |
| CFB 5 | 0.24 |
| CFB 10 | 0.15 |
| CFB 5 W 1 | 0.13 |
| CFB 5 W 7 | 0.2 |
| CFB 5 W 10 | 0.2 |

The background-subtracted GI-XRD spectra for Si substrate, Si/CFB ($t_{CFB}$), Si/W ($t_W$), and Si/CFB ($t_{CFB}$ = 5 nm)/W ($t_W$) thin films are presented in Fig. 3. Peak-to-baseline intensities are sustained as obtained during the XRD measurement in these plots. For bare Si substrate, we have obtained a peak near $52^0$, which corresponds to Si (211) in GI mode[29,30]. The peak is present consistently for most of the magnetic samples. The bulk XRD from the same substrate confirms the presence of Si (100) peak near $69^0$ (Fig. 4S(f)).

The XRD spectra for Si/CFB ($t_{CFB}$) films are shown in Fig. 3(a). The absence of a crystalline peak in the spectrum is indicative of the predominantly amorphous structure of Si/CFB ($t_{CFB}$) films. B transforms crystalline CoFe into an amorphous alloy by inducing atomic-scale disorder and preventing crystal formation [26,27].

The XRD spectra for bare Si/W ($t_W$) films substrate are presented in Fig. 3(b). Development of prominent Bragg diffraction peaks are observed at $2\theta$ =36.12°, 43.77°, 58.00°, 67.51°, 69.88°and 73.08° which correspond to the $\beta$-(200), $\beta$-(211), $\alpha$-(200), $\beta$-(320), $\beta$-(321), and $\alpha$-(211) planes, respectively. A high intensity peak is observed at $2\theta = 40.00°$ which is the superimposed peaks of the $\beta$-(210) and $\alpha$-(110) planes since these two peaks are very close to each other at $\Delta 2\theta \cong 0.30°$. Usually, $\beta$-(210) and $\alpha$-(110) exist in the close vicinity of 40°. The observations matched with the reference Crystallographic open database and supported by findings reported in the literatures [31,32]. For films with thicknesses of 3 nm and above, the peaks at 36.12°, 40.13°, and 43.77° became clearly detectable, indicating the initiation of crystallization. For films with thicknesses of 7 nm and 10 nm, additional peaks at 67.51°, 69.88°, and 73.08° emerged, and one additional peak appeared at 58° for 10 nm, signifying further crystallite growth and improved phase development with increasing film thickness. For $t_W$ > 10 nm, and additional α-peak near 58° confirms the development of **α**-dominated phase. The XRD spectra for the Si/CFB ($t_{CFB}$ = 5 nm)/W ($t_W$) samples are shown in Fig. 3(c). The Bragg diffraction peaks corroborate with the peaks achieved for bare Si/W ($t_W$) films, as the XRD intensity reflects the crystallinity of the top W layer.

It was reported earlier for such materials; the ordering could be observed above 8 nm. $\alpha + \beta$ phase shows that the Si/W ($t_W$) thin films are in a metallic state. The transition thickness from β to $\alpha + \beta$ phase may be increased or decreased based on the deposition conditions of Si/W ($t_W$) films, which is reported in earlier studies [33,34]. The underlined phenomena behind the growth of α+ β phase is the moderate growth rate with sufficient kinetic energy and surface mobility at room temperature [31]. The confirmation of $\alpha + \beta$ phase above 7 nm of Si/W ($t_W$) is confirmed by electrical measurement in the Van der Pauw geometry by using custom built four-probe setup, where the electrical resistivity is $\approx 100\ \mu\Omega$ cm, corroborates with the value from existing literature [35].

To analyze the XRD spectra, we have fitted the experimental data with Gaussian peak functions. The FWHM extracted from this fitting ranges between 1.16° and 0.25°. Now applying Debye-Scherrer formula, crystallite size is calculated from individual peaks which range between 3.20 to 10.71 nm for Si/W ($t_W$) films and 7.51 to 20.70 nm for Si/CFB ($t_{CFB}$ = 5 nm)/W ($t_W$) layers. Fig. 4S contains detailed variation of these parameters.

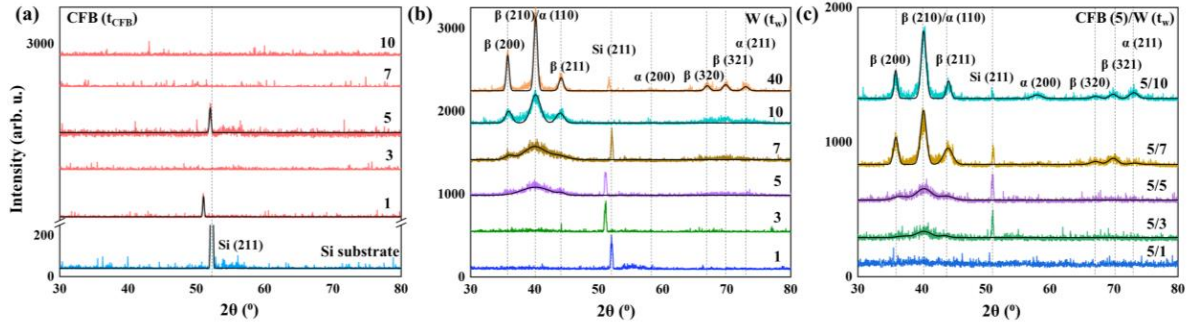

Figure 3: The XRD spectra for (a) Si/CFB ($t_{CFB}$), along with Si substrate (b) Si/W ($t_W$) and Si/CFB ($t_{CFB}$ = 5 nm)/W ($t_W$) obtained in GI mode. The numbers indicate thickness in nm. The (hkl) values are shown near the respective peaks. The $\alpha$ and $\beta$ correspond to the different phases of the W. Solid black curves represent the fitted peaks with Gaussian peak function.

### 3.2 Room temperature static magnetic properties

Figure 4(a-b) represents the normalized hysteresis loops obtained in IP and OOP configurations for Si/CFB ($t_{CFB}$) films with varying $t_{CFB}$. Fig. 3(c, d) shows the same for Si/CFB ($t_{CFB}$ = 5 nm)/W ($t_W$) films with different $t_W$. Ultrathin a-CFB films having a thickness ranging between 0.7 and 3.0 nm exhibit tilted loops for both configurations. The superposition of IP and OOP loops with $H_s \approx$ 8700 Oe, indicates a robust tilt in the magnetic layer towards the OOP direction from the film plane. As thickness increases, the IP loops appear to be stiffer in comparison to the OOP loop. This indicates a preference for the magnetization to lie in the plane. The magnetization reversal is gradual to the applied magnetic field. This is designated as an interface-anisotropy-dominated (IAD) regime. A detailed justification is provided later in this paper.

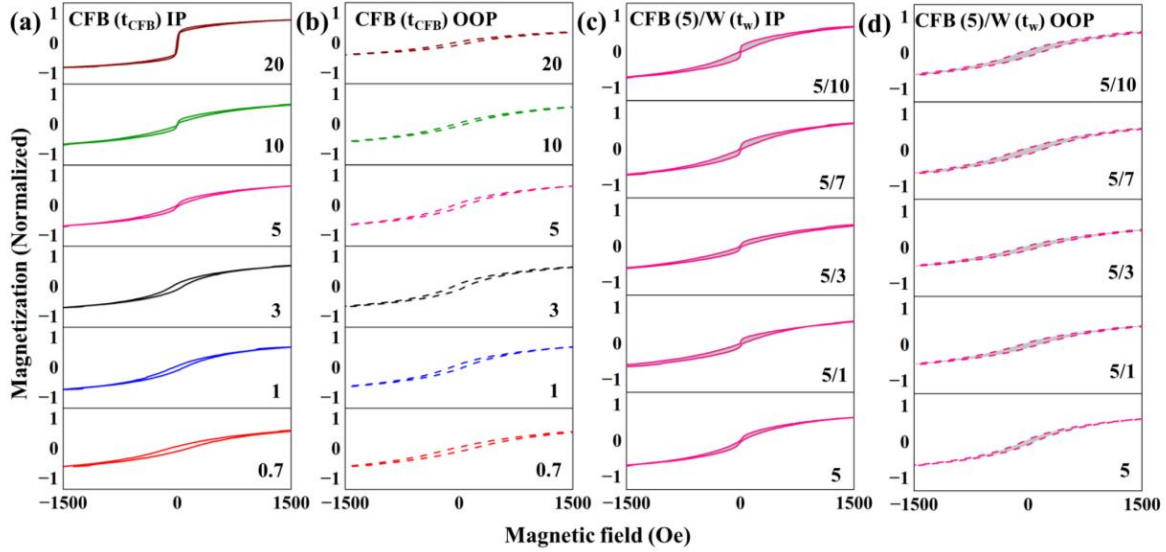

Figure 4: (a-d) Normalized hysteresis loops for Si/CFB ($t_{CFB}$) and Si/CFB ($t_{CFB}$ = 5 nm)/W ($t_W$) films. IP loops (solid line) for Si/CFB ($t_{CFB}$) and Si/CFB ($t_{CFB}$ = 5 nm)/W ($t_W$) films are shown in (a) and (c) respectively in the range ± 1500 Oe. The OOP loops (dashed line) for Si/CFB ($t_{CFB}$) and Si/CFB ($t_{CFB}$ = 5 nm)/W ($t_W$) films are shown in (b) and (d) in the same range. The numbers inside the plots represent the thickness in nm.

The IP loops for $t_{CFB} \geq 5$ nm exhibit anomalous behavior with two-stage reversal, and for $t_{CFB} = 10$ nm, the loop is completely inverted. As the applied field is swept between the positive and negative saturation values, the magnetization reverses at $H \approx 2$ Oe. This inverted loop with negative remanence has been rarely observed in a-CFB films. We have defined this as a growth-induced shape anisotropy-dominated (GI-SAD) regime. With increasing film thickness to 20 nm, the IP loop shape has a conventional 'nearly rectangular' feature. The OOP loops are more tilted than the IP loop, indicating a transition towards IP magnetic configuration. This is conventional shape anisotropy dominated (SAD) regime.

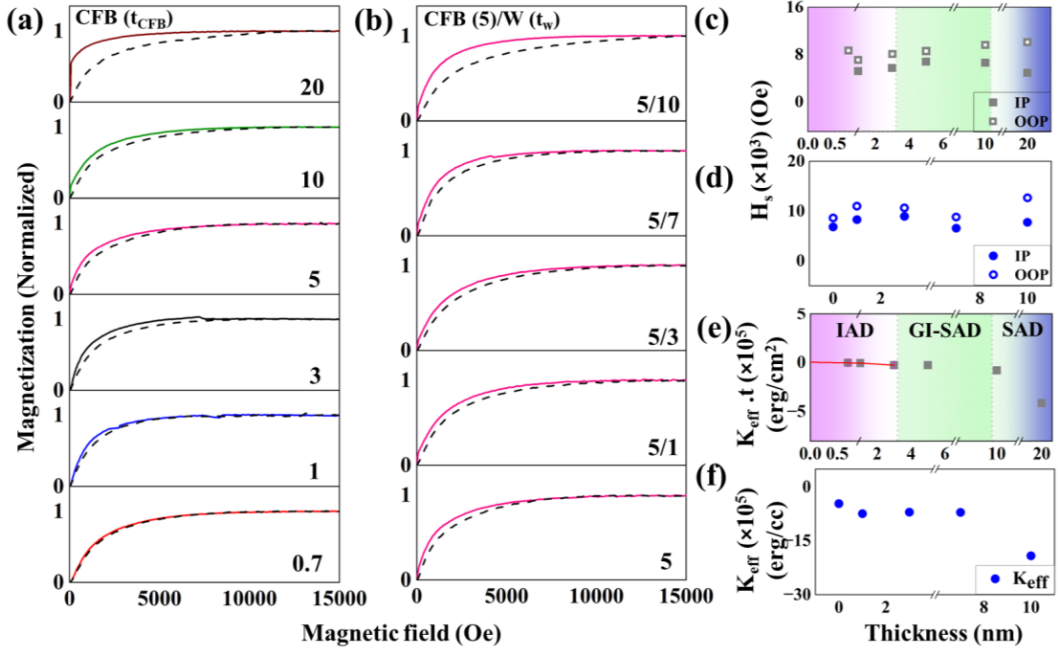

Figure 5: Normalized IP (solid line) and OOP (dashed line) initial curves for (a) Si/CFB ($t_{CFB}$) and (b) Si/CFB ($t_{CFB}$ = 5 nm)/W ($t_W$) films, respectively. Evolution of $H_s$ with thickness for (c) Si/CFB ($t_{CFB}$) and (d) Si/CFB ($t_{CFB}$ = 5 nm)/W ($t_W$). (e) Variation of $K_{eff} \cdot t_{CFB}$ with thickness ($t_{CFB}$) for Si/CFB ($t_{CFB}$) and (f) variation of $K_{eff}$ for Si/CFB ($t_{CFB}$ = 5 nm)/W ($t_W$) with thickness ($t_W$). The solid red line represents the fitting of $K_{eff} \cdot t_{CFB}$ values. The numbers inside the plots represent the thickness in nm.

Figure 5(a-b) shows the normalized initial curves for Si/CFB ($t_{CFB}$) and Si/CFB ($t_{CFB}$ = 5 nm)/W ($t_W$) samples obtained in the IP and OOP configurations. The variation of $H_s$ (IP) and $H_s$ (OOP) are plotted with increasing $t_{CFB}$. For $t_{CFB}$ = 0.7 nm, the values are comparable. This indicates that the domains can be oriented in these two directions by the expense of similar energy provided by the external magnetic field. This in turn validates that fact that there is a robust tilt of ≈45° in the magnetization of the film. As the film thickness increases, the difference between $H_s$ (IP) and $H_s$ (OOP) becomes prominent. The magnetization of the films gradually aligns itself in the plane of the film. Following the area method, the difference in area between the IP and OOP curves are converted into anisotropy energy constant ($K_{eff}$). The variation of $K_{eff}$ with $t_{CFB}$ is nonmonotonic in nature. The $K_{eff}$ has two major contributors: volume anisotropy and surface anisotropy. In the IAD regime, the Si/CFB ($t_{CFB}$) interface gives rise to Neel-type anisotropy ($K_N$) [36]. Despite the presence of volume anisotropy ($K_V$) originating from the shape of the thin film ($K_{SA}$) and anisotropies of the bulks atoms ($K_B$), the interface effects prevail for $t_{CFB} \leq$ 3 nm. The PMA energy

is calculated by using the area method (detailed discussion can be found in the supplementary material) can be expressed by the following equation:

$$K_{eff} = K_V + K_S/t_{CFB} = K_B - K_{SA} + K_S/t_{CFB} \quad (1a)$$

$$K_{eff} \cdot t_{CFB} = (K_B - K_{SA}) \cdot t_{CFB} + K_S \quad (1b)$$

After fitting the data points (see Fig. 5(c)) with this equation 1b till $t_{CFB} \leq 3$ nm, we obtained the slope, $K_V = 1 \times 10^6$ erg/cc and intercept, $K_S = K_N = 0.06$ erg/cm$^2$. The negative value of the slope represents that the films exhibit progress from OOP to IP magnetization in the demagnetized state. The $K_S$ is independent of thickness, and this reportedly low anisotropy [36] contributes to all the films; however, it becomes dormant in the thicker regime due to the presence of other bulk contributors that scale strongly with thickness. The estimated values for $K_B$ and $K_{SA}$ are $9 \times 10^6$ erg/cc and $8 \times 10^6$ erg/cc, respectively. According to the existing reports, additional anisotropy can originate from the bulk atoms in films having a composite-type of magnetic environment and uniformly distributed magnetic inhomogeneities. Additionally, these films are uncapped, and the possibility of forming an ultrathin self-oxidized top layer cannot be ruled out. This might affect magnetic homogeneity. The dominance of several interface effects leading to interface anisotropy is studied extensively in the ultrathin CFB films: spin-orbit interactions at the interface [37], strain induced at the interface [38], thickness gradient in the film growth [39], due to inhomogeneous wetting or island formation [40] etc. We anticipate that the weak spin-orbit interaction between the magnetic layer and the self-oxidized silicon substrate can result in a canting of the spins in the ferromagnetic layer, which ultimately gives rise to the interface anisotropy. Moreover, the applied mechanical stress or stress induced due to growth conditions, lattice mismatch etc. might induce strain between the FM layer and substrate. However, we rule out the presence of such contributors in these films. The growth conditions used for the deposition of these films can also lead to the formation of a thickness gradient in the ultrathin regime $t_{CFB} \leq 3$ nm [39]. As CFB has higher surface energy than the Si substrate, the tendency of incoming atoms from the CFB target to adhere to the atom of the same kind, becomes pronounced. This supports the formation and retention of isolated clusters during the initial film growth, which can be explained by Volmer-Weber growth model [41]. These clusters of inhomogeneity will be produced due to the insufficiency of the ad-atoms to coalesce as these atoms are cooled and solidified instantaneously, leaving no time to meld uniformly within the film. The inhomogeneities created in the film expedite the pinning of magnetization in certain directions. Nonetheless, growth-induced stress contribution to magnetic anisotropy will be negligible in this thickness regime. Instead, nm-thick film might not have magnetic continuum and instead can be taken as a collection of discrete magnetic dipoles. The dipolar interaction of outer and inner layers can also differ, leading to the volume as well as surface contribution of this anisotropy. However, the interface dipolar contribution is of lesser significance than the other interface anisotropy contributors. The competition between the conventional SA (volume anisotropy) and the above-mentioned anisotropies (surface anisotropies) causes a reasonable tilt of magnetization in this regime. As the film thickens up to 3 nm, it becomes more continuous, hence, contains lesser inhomogeneities, and a significant SA contribution due to less discontinuity in the film results in the mutation of tilt towards the film plane. In this thickness range ($t_{CFB} > 3$ nm) the atoms approaching the substrate sit over the islands following the direction of the flux, resulting in a shadowing effect, causing column-like structures to form. The canted columns can have different shapes depending on the growth conditions. Initially, the columns might be of rod-shaped, tomb-shaped, pyramid shape etc. [42]. The magnetization will point towards the long axis of the column. However, if the OOP canting is not significant, then a sizable in-plane projection of the canted magnetization may result in the bulk UMA. See supplementary information for more detailed analysis. This is designated by growth-induced shape anisotropy (GI-SA). It is pertinent to mention that the relative angle between this UMA and the applied magnetic field during hysteresis is assumed to be small, however, it could not be quantified in the present work. The films can also be subjected to internal stress due to growth conditions. This stress can significantly modify the magnetization orientation for the magnetostrictive material. The magnetization direction will be determined by the combined effect of the sign of the stress ($\sigma$) and that of the magnetostriction constant ($\lambda$) [43]. The nature of the GI-stress (planar tensile or planar compressive) is decided by a few major technical factors: deposition power, Ar pressure, deposition geometry, and heat treatment etc. [44]. Low power and moderate Ar pressure can lead to compressive stress. The formation of dense columnar structures leads to compressive stress, and under-dense structures lead to tensile stress, depending on the deposition geometry [45]. Detailed discussion about the effect of heat treatment can be found elsewhere. The magnetization prefers the direction of stress if $\lambda\sigma > 0$. The competition between all these individual anisotropies leads to effective anisotropy that will ultimately decide the easy and hard directions for the magnetization in a film. This $K_{eff}$, due to the combined contribution of these anisotropies, can be written as:

$$K_{eff} = K_{SA} + K_{GI-SA} + K_{MsA} + ……  \quad (2)$$

The nature of $\lambda$ in the case of our a-CFB films is identified to be positive from the composition of the films. Combined with the positive $\lambda$ of this set of a-CFB films, forces magnetization to lie out of the sample plane. The conventional SA is independent of thickness, and it appears to be the major contributor in the SAD regime. We anticipate that the contribution from GI-SA weakens because of the decreased shadowing effect due to substrate rotation (though the speed is slow) [46]. It implies dominance of IP anisotropy. It is typically observed that stress-induced anisotropies become stronger with film thickness beyond our experimental thickness range.

It is extremely important to discuss here that the addition of boron can lead to the distortion of the α-iron structure. This is the primary cause of the formation of amorphous or nanocrystalline structures of Fe-based materials, as segregation of boron is reported on the grain boundary [47]. Diffusion of this element may improve the TMR of the system [48]. Relatively high-boron content can modify the magnetic properties of the host alloy. It may increase the effective anisotropy field while decreasing coercivity. However, it has a negligible influence on the modification of the effective magnetic moment of the sample. [47] Apart from the magnetic properties, modification of electrical properties (*i.e.* enhancement of resistivity) is reported in alloys with high-boron concentration. A slight increase in Boron concentration also leads to a reduction of positive magnetostriction of CFB alloys. We suspect a modification of composition in these sputter-deposited films with respect to the target, which is previously reported in other studies [2]. However, quantification of Boron-mediated changes in the magnetic properties and improvement of the same is beyond the scope of this paper. We believe the overall trend in thickness-dependent variation of the material properties remains unaffected by this.

The 5-nm-thick CFB film is capped with a W layer of different thicknesses. As discussed in the previous section, the W is having a mixed-α-β phase while seeded on CFB layer. W is usually a good capping agent. We observed that the primary trend of the magnetization reversal for these films corroborates 5 nm bare film. However, the saturation field and effective anisotropy values increase for the sample with 10-nm-thick W on the top. As we mentioned earlier, the uncapped films might have suffered from self-oxidation; capping improves the magnetic homogeneity of the CFB layer adjacent to the W layer. The uniformity of the capping layer also improves with thickness, and it counters the effect from OOP contributors competing with the inherent shape anisotropy of the film.

## 4. Conclusion

Technologically important ultrathin-to-thin a-CFB films, deposited by rf-magnetron sputtering on a self-oxidized Si (100) substrate, are characterized by several techniques. The surface morphology of both uncapped and capped films reveals that the surfaces are smooth and robust against environmental degradation. The GI-XRD spectra for the bare CFB films contain only the substrate peak. The series of CFB films, ranging between 0.7 and 20 nm, is amorphous in nature. The GI-XRD data obtained from the W capping layer confirms the presence of mixed-$\alpha - \beta$ phases. Magnetic hysteresis behavior is studied by using a VSM system in the IP and OOP configurations. The shape of the IP loops evolving from an 'upright S' to 'nearly rectangular' nature *via* a 'completely inverted' profile indicated thickness-dependent emergence of magnetic anisotropies. It is important to mention that a self-sustained tilted magnetic anisotropy is stabilized in a seed-free environment based on the direct substrate-to-magnet interaction. This avoids ambiguities of the multistep metallization process. Qualitative deconvolution of complex anisotropic behavior reveals that the interplay between interface anisotropy, conventional shape anisotropy, growth-induced anisotropies, and magnetic inhomogeneity-induced anisotropies competes to counter each other. Based on their dominance, we identify three anisotropy regimes: IAD, GI-SAD, and SAD. Interfacial anisotropy contribution is also quantified for the IAD regime. Our work provides novel insight that might be beneficial for designing energy-efficient spintronic devices based on amorphous soft magnetic alloys.

## Acknowledgements

We acknowledge the financial support from Shiv Nadar Institution of Eminence (Delhi NCR) and INSPIRE fellowship (Department of Science and Technology, Govt. of India, grant number: G-CONF000220). K.K., B.G., and A.M. acknowledge Shiv Nadar Institution of Eminence (Delhi NCR) for research fellowships. We would also like to acknowledge the financial support from the Department of Science and Technology, India, under the FIST project (grant number: SR/FST/PS-I/2017/6(C)) for the installation and maintenance of a custom-built Hall measurement setup. We are grateful to our scientific officers, Dr. Rakesh Kumar, Dr. Arpan Bhattacharya, and Dr. Khushboo Agarwal from the Department of Physics (Shiv Nadar Institution of Eminence, Delhi NCR), for their guidance in handling PPMS, sputtering technique, and resistivity measurements. We are glad to acknowledge technical support from Mr. Ravinder Singh and Dr. Aniruddha Das from the Department of Chemistry (Shiv Nadar Institution of Eminence, Delhi NCR) for AFM and XRD (Rigaku X-ray diffractometer 3kW system, model Smartlab SE funded by Shiv Nadar Foundation) measurements, respectively. We also acknowledge Dr. Raju Vemoori from the Department of Mechanical Engineering (Shiv Nadar Institution of Eminence, Delhi NCR) for helping us obtain EDX data. We gratefully acknowledge the intellectual support and guidance received from Dr. Jaivardhan Sinha (SRM-Chennai) during the optimization of our alloy deposition.

## Conflict of interest

The authors have declared no conflict of interest.

## Keywords

Amorphous alloy, Crystalline heavy-metal, Magnetic anisotropy, Ultra-thin film, Magnetic hysteresis, Vibrating Sample Magnetometry

# References

[1] N. Heiman, R.D. Hempstead, N. Kazama, Low-coercivity amorphous magnetic alloy films, J Appl Phys 49 (1978) 5663–5667. https://doi.org/10.1063/1.324489.

[2] A. Brunsch, Non-magnetostrictive amorphous soft magnetic thin film material of high magnetization, J Appl Phys 50 (1979) 7600–7602. https://doi.org/10.1063/1.326857.

[3] S. Tsunashima, Y. Maehata, S. Uchiyama, Induced anisotropy and permeability in amorphous Fe-B and Co-Fe-B films, IEEE Trans Magn 17 (1981) 3073–3075. https://doi.org/10.1109/TMAG.1981.1061583.

[4] U. Dibbern, Magnetic Field Sensors Using The Magnetoresistive Effect, 1986.

[5] Kim Dae Yong, A study of the effect of process-parameters on the magnetic anisotropy of Cobalt-based soft amorphous thin films, The University of Texas at Austin, 1988.

[6] T. Uchiyama, K. Mohri, L. V Paninaand, K. Furuno, Magneto-Impedance in Sputtered Amorphous Films for Micro Magnetic Sensor, 1995.

[7] K.K.H.M.S.T. and S.U. Mutsuko Jimbo, Giant Magnetoresistance Effect in Amorphous CoFeB Sandwiches, Jpn J Appl Phys 34 (1995) 112.

[8] M. Jimbo, K. Komiyama, S. Tsunashima, Giant magnetoresistance effect and electric conduction in amorphous-CoFeB/Cu/Co sandwiches, J Appl Phys 79 (1996) 6237–6239. https://doi.org/10.1063/1.362016.

[9] M.J.Y.I.K.K. S. Tsunashima, Spin valves using amorphous magnetic layers, J Magn Magn Mater 165 (1997) 111–114.

[10] T. Feng, J.R. Childress, Fabrication of exchange-biased spin valves with CoFeB amorphous layers, J Appl Phys 85 (1999) 4937–4939. https://doi.org/10.1063/1.370051.

[11] R.B.G.K.-J.L. Bernard Dieny, Introduction to Magnetic Random-Access Memory, Wiley–IEEE Press, 2016.

[12] Y. Shirota, S. Tsunashima, R. Imada, Y. Nomura, S. Iwata, M. Jimbo, Giant Magnetoresistance Effect in CoFeB/Cu/CoFeB Spin Valves, Publication Board, 1999. http://iopscience.iop.org/1347-4065/38/2R/714.

[13] H.J. Kim, S.C. Oh, J.S. Bae, K.T. Nam, J.E. Lee, S.O. Park, H.S. Kim, N.I. Lee, U.I. Chung, J.T. Moon, H.K. Kang, Development of magnetic tunnel junction for toggle MRAM, IEEE Trans Magn 41 (2005) 2661–2663. https://doi.org/10.1109/TMAG.2005.854935.

[14] R.W. Dave, G. Steiner, J.M. Slaughter, J.J. Sun, B. Craigo, S. Pietambaram, K. Smith, G. Grynkewich, M. DeHerrera, J. Åkerman, S. Tehrani, MgO-based tunnel junction material for high-speed toggle magnetic random access memory, IEEE Trans Magn 42 (2006) 1935–1939. https://doi.org/10.1109/TMAG.2006.877743.

[15] J.U. Cho, D.K. Kim, T.X. Wang, S. Isogami, M. Tsunoda, M. Takahashi, Y.K. Kim, Magnetoresistance variation of magnetic tunnel junctions with NiFeSiB/CoFeB free layers depending on MgO tunnel barrier thickness, IEEE Trans Magn 44 (2008) 2547–2550. https://doi.org/10.1109/TMAG.2008.2003244.

[16] A. Kaidatzis, V. Psycharis, J.M. Garcia-Martin, C. Bran, M. Vazquez, D. Niarchos, Perpendicular magnetic anisotropy on W-based spin-orbit torque CoFeB | MgO MRAM Stacks, in: Materials Research Society Symposium Proceedings, Materials Research Society, 2015: pp. 73–78. https://doi.org/10.1557/opl.2015.191.

[17] K. Fritz, L. Neumann, M. Meinert, Ultralow switching-current density in all-amorphous W-Hf/Co-Fe-B/Ta Ox films, Phys Rev Appl 14 (2020). https://doi.org/10.1103/PhysRevApplied.14.034047.

[18] Y.Z.Y.S.Z.L.J.Z.H.D.X.Y.Q.J.Y.R.Z.Z. and Z.Z. X. Ma, THz emission manipulation and thermal robustness in CoFeB films via boron doping, Appl. Phys. Lett. 127 (2025).

[19] X. Han, X. Yao, T.R. Thapaliya, G. Bierhance, C. In, Z. Ni, A. Bedoya-Pinto, S. Huang, C. Felser, S.S.P. Parkin, T. Kampfrath, S. Oh, L. Wu, Broad-band THz emission by Spin-to-Charge Conversion in Topological Material -- Ferromagnet Heterostructures, (2025). http://arxiv.org/abs/2507.14838.

[20] S.U. Jen, Y.D. Yao, Y.T. Chen, J.M. Wu, C.C. Lee, T.L. Tsai, Y.C. Chang, Magnetic and electrical properties of amorphous CoFeB films, J Appl Phys 99 (2006). https://doi.org/10.1063/1.2174113.

[21] S.X. Huang, T.Y. Chen, C.L. Chien, Spin polarization of amorphous CoFeB determined by point-contact Andreev reflection, Appl Phys Lett 92 (2008). https://doi.org/10.1063/1.2949740.

[22] J.M. Teixeira, R.F.A. Silva, J. Ventura, A.M. Pereira, F. Carpinteiro, J.P. Araújo, J.B. Sousa, S. Cardoso, R. Ferreira, P.P. Freitas, Domain imaging, MOKE and magnetoresistance studies of CoFeB films for MRAM applications, Materials Science and Engineering: B 126 (2006) 180–186. https://doi.org/10.1016/j.mseb.2005.09.031.

[23] S. Ikeda, K. Miura, H. Yamamoto, K. Mizunuma, H.D. Gan, M. Endo, S. Kanai, J. Hayakawa, F. Matsukura, H. Ohno, A perpendicular-anisotropy CoFeB-MgO magnetic tunnel junction, Nat Mater 9 (2010) 721–724. https://doi.org/10.1038/nmat2804.

[24] A.T. Hindmarch, A.W. Rushforth, R.P. Campion, C.H. Marrows, B.L. Gallagher, Origin of in-plane uniaxial magnetic anisotropy in CoFeB amorphous ferromagnetic thin films, Phys Rev B Condens Matter Mater Phys 83 (2011). https://doi.org/10.1103/PhysRevB.83.212404.

[25] L. Kipgen, H. Fulara, M. Raju, S. Chaudhary, In-plane magnetic anisotropy and coercive field dependence upon thickness of CoFeB, J Magn Magn Mater 324 (2012) 3118–3121. https://doi.org/10.1016/j.jmmm.2012.05.012.

[26] A. Gayen, G.K. Prasad, S. Mallik, S. Bedanta, A. Perumal, Effects of composition, thickness and temperature on the magnetic properties of amorphous CoFeB thin films, J Alloys Compd 694 (2017) 823–832. https://doi.org/10.1016/j.jallcom.2016.10.066.

[27] P. Kumar, V. Sharma, M.K. Khanna, B.K. Kuanr, Effect of ferromagnetic layer thickness on the static and dynamic magnetic properties of sputter grown thin CoFeB films, Physics Letters, Section A: General, Atomic and Solid State Physics 553 (2025). https://doi.org/10.1016/j.physleta.2025.130722.

[28] R. Lavrijsen, A. Feránndez-Pacheco, D. Petit, R. Mansell, J.H. Lee, R.P. Cowburn, Tuning the interlayer exchange coupling between single perpendicularly magnetized CoFeB layers, Appl Phys Lett 100 (2012). https://doi.org/10.1063/1.3682103.

[29] A. Sahoo, S. Mallick, A. Rath, H. Ding, A. Azevedo, S. Bedanta, Efficient control of magnetization dynamics via W/CuOX interface, Appl Phys Lett 125 (2024). https://doi.org/10.1063/5.0230730.

[30] A. Sahoo, S.P. Mahanta, S. Bedanta, Significant influence of low SOC materials on magnetization dynamics and spin-orbital to charge conversion, Npj Spintronics 3 (2025). https://doi.org/10.1038/s44306-025-00080-5.

[31] K. Sriram, R. Mondal, J. Pradhan, A. Haldar, C. Murapaka, Structural Phase Engineering of (α + β)-W for a Large Spin Hall Angle and Spin Diffusion Length, Journal of Physical Chemistry C 127 (2023) 22704–22712. https://doi.org/10.1021/acs.jpcc.3c04404.

[32] J.-S. Lee, J. Cho, C.-Y. You, Growth and characterization of α and β -phase tungsten films on various substrates , Journal of Vacuum Science & Technology A: Vacuum, Surfaces, and Films 34 (2016). https://doi.org/10.1116/1.4936261.

[33] Q. Hao, W. Chen, G. Xiao, Beta (β) tungsten thin films: Structure, electron transport, and giant spin Hall effect, Appl Phys Lett 106 (2015). https://doi.org/10.1063/1.4919867.

[34] J. Yu, X. Qiu, W. Legrand, H. Yang, Large spin-orbit torques in Pt/Co-Ni/W heterostructures, Appl Phys Lett 109 (2016). https://doi.org/10.1063/1.4959958.

[35] S. Mondal, S. Choudhury, N. Jha, A. Ganguly, J. Sinha, A. Barman, All-optical detection of the spin Hall angle in W/CoFeB/SiO2 heterostructures with varying thickness of the tungsten layer, Phys Rev B 96 (2017). https://doi.org/10.1103/PhysRevB.96.054414.

[36] M.T. Johnson, J.H. Bloemen, J.A. Den Broeder, J.J. De Vries, Magnetic anisotropy in metallic multilayers, 1996.

[37] A.S. Silva, S.P. Sá, S.A. Bunyaev, C. Garcia, I.J. Sola, G.N. Kakazei, H. Crespo, D. Navas, Dynamical behaviour of ultrathin [CoFeB (tCoFeB)/Pd] films with perpendicular magnetic anisotropy, Sci Rep 11 (2021). https://doi.org/10.1038/s41598-020-79632-0.

[38] G. Yu, Z. Wang, M. Abolfath-Beygi, C. He, X. Li, K.L. Wong, P. Nordeen, H. Wu, G.P. Carman, X. Han, I.A. Alhomoudi, P.K. Amiri, K.L. Wang, Strain-induced modulation of perpendicular magnetic anisotropy in Ta/CoFeB/MgO structures investigated by ferromagnetic resonance, Appl Phys Lett 106 (2015). https://doi.org/10.1063/1.4907677.

[39] P. Vineeth Mohanan, K.R. Ganesh, P.S. Anil Kumar, Spin Hall effect mediated current-induced deterministic switching in all-metallic perpendicularly magnetized Pt/Co/Pt trilayers, Phys Rev B 96 (2017). https://doi.org/10.1103/PhysRevB.96.104412.

[40] Samridh Jaiswal, Investigation of the Dzyaloshinskii-Moriya interaction and perpendicular magnetic anisotropy in magnetic thin films and nanowires, 2018.

[41] C.I. Fornari, G. Fornari, P.H. de O. Rappl, E. Abramof, J. dos S. Travelho, Monte Carlo Simulation of Epitaxial Growth, in: Epitaxy, InTech, 2018. https://doi.org/10.5772/intechopen.70220.

[42] A. Barranco, A. Borras, A.R. Gonzalez-Elipe, A. Palmero, Perspectives on oblique angle deposition of thin films: From fundamentals to devices, Prog Mater Sci 76 (2016) 59–153. https://doi.org/10.1016/j.pmatsci.2015.06.003.

[43] C.D.G. B. D. Cullity, Introduction to Magnetic Materials, 2nd ed., Wiley-IEEE Press, 2009.

[44] S.F. Cheng, P. Lubitz, Y. Zheng, A.S. Edelstein, Effects of spacer layer on growth, stress and magnetic properties of sputtered permalloy film, in: J Magn Magn Mater, 2004: pp. 109–114. https://doi.org/10.1016/j.jmmm.2004.04.027.

[45] J.A. Thornton, J. Tabock, D.W. Hoffman, Internal Stresses In Metallic Films Deposited By Cylindrical Magnetron Sputtering*, 1979.

[46] Y. Shim, M.E. Mills, V. Borovikov, J.G. Amar, Effects of substrate rotation in oblique-incidence metal(100) epitaxial growth, Phys Rev E Stat Nonlin Soft Matter Phys 79 (2009). https://doi.org/10.1103/PhysRevE.79.051604.

[47] I. Kim, J. Kim, K.H. Kim, M. Yamaguchi, Effects of boron contents on magnetic properties of Fe-Co-B thin films, IEEE Trans Magn 40 (2004) 2706–2708. https://doi.org/10.1109/TMAG.2004.832129.

[48] S. Mukherjee, R. Knut, S.M. Mohseni, T.N. Anh Nguyen, S. Chung, Q. Tuan Le, J. Åkerman, J. Persson, A. Sahoo, A. Hazarika, B. Pal, S. Thiess, M. Gorgoi, P.S. Anil Kumar, W. Drube, O. Karis, D.D. Sarma, Role of boron diffusion in CoFeB/MgO magnetic tunnel junctions, Phys Rev B Condens Matter Mater Phys 91 (2015). https://doi.org/10.1103/PhysRevB.91.085311.

# Supplementary Information

**Estimation of deposition rate and film thickness**

Figure 1S (a-e) contains the height profile from the line scans procured using a stylus-based profilometer [Bruker Dektak profilometer]. The thickness values for $Co_{20}Fe_{60}B_{20}$ (CFB) and Tungsten (W) films may vary by 4-5% from the nominal thickness range. The deposition rates are estimated to be 0.6 Å/sec. and 0.4 Å/sec., respectively for CFB and W films as obtained from the linear fit of thickness vs deposition time (see Fig. 1S(f)). The lower thicknesses are obtained by extrapolating these values.

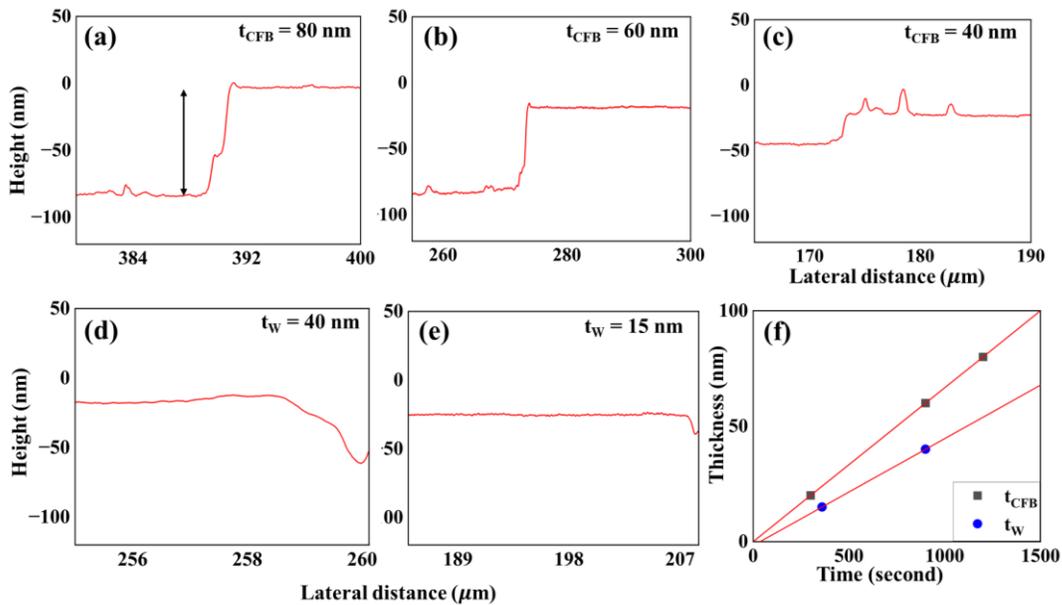

Figure 1S: Line-scan thickness profile for (a-c) Si/CFB ($t_{CFB}$ = 80, 60 and 40 nm), (d-e) Si/W ($t_W$ = 40 and 15 nm). The double-sided arrow represents the step height. (f) Thicknesses vs deposition time. The linear fits to estimate deposition rate are shown for both materials.

**AFM images for rest of the samples**

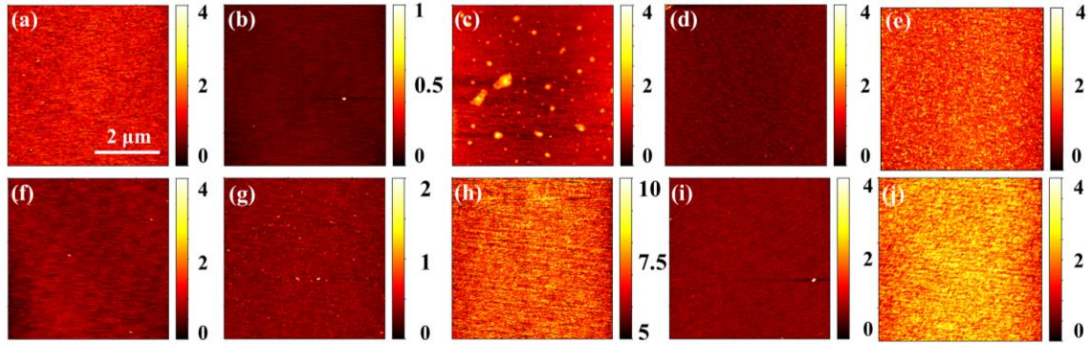

Figure 2S: AFM images of (a-c) Si/CFB ($t_{CFB}$ = 0.7, 3, and 7 nm), (d-e) Si/CFB ($t_{CFB}$ = 5 nm)/ W ($t_W$ = 3 and 5 nm), and (f-i) Si/W ($t_W$ = 1, 3, 5, 7, and 10 nm) processed using Gwyddion software. The scan area is of 5μm × 5 μm. The length scale is given for reference. Height profiles are depicted by colour bars.

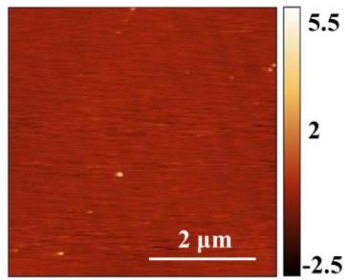

Figure 3S: AFM image of Si substrate processed using Gwyddion software. The scan area is of 5μm × 5 μm. The length scale is given for reference. Height profiles are depicted by colour bars.

## XRD analysis for samples and substrate

Figure 4S shows the parameters obtained from GI-XRD spectra for both Si/W ($t_W$) and Si/CFB ($t_{CFB}$ = 5 nm)/W ($t_W$) and XRD spectra for the Si substrate in bulk mode. The FWHM is obtained by fitting the peaks with a Gaussian peak function. Debye-Scherrer formula has been employed to calculate crystallite size [1]:

$$D = \frac{k\lambda'}{\xi \cos\theta}$$

where $D$ is crystallite size, $k$ is the shape factor (0.9), $\lambda'$ is the x-ray wavelength, $\xi$ is the line broadening at half the maximum intensity (FWHM), and $\theta$ is the Bragg's angle. The crystallite sizes of Si/CFB ($t_{CFB}$ = 5 nm)/W ($t_W$) are observed to be higher than those of Si/W ($t_W$). The (100) orientation of the Si substrate is verified by the presence of the (400) peak at 69.3° from the XRD spectra of the Si substrate in bulk mode.

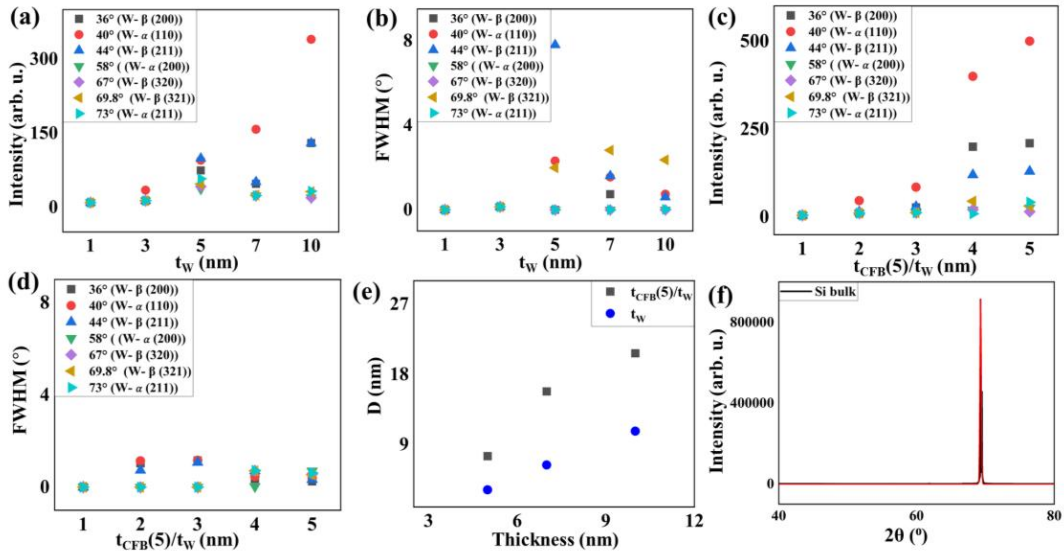

Figure 4S: Variation of intensity and FWHM for (a-b) Si/W ($t_W$) and (c-d) Si/CFB ($t_{CFB}$ = 5 nm)/W ($t_W$) respectively. (e) crystallite sizes ($D$) for both Si/W ($t_W$) and Si/CFB ($t_{CFB}$ = 5 nm)/W ($t_W$) and (f) XRD spectra for the Si substrate in bulk mode.

# Effect of the oblique angle deposition on the structure and the mechanical properties of the film

Thin films deposited with certain oblique angles through PVD techniques such as magnetron sputtering can result in tilted nanopillars or column-like structures. These columns can have different shapes and tilts, which depend on the deposition geometry. The following formula can be used to calculate the angle with which these columns are slanted with respect to the substrate normal ($\beta'$):

$$tan(\alpha') = 2tan(\beta')$$

Where, $\alpha'$ is angle of deposition, the angle between incoming flux and substrate normal [2] (see Fig 5S) Since, in our case, $\alpha' \approx 45°$, which leads to $\beta' \approx 22°$. The formation of such a structure can result in the origin of growth-induced shape anisotropy (GI-SA).

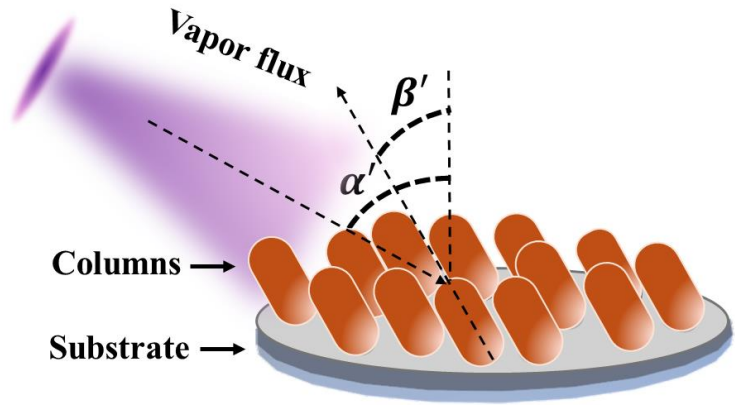

Figure 5S: Pictorial representation of possible columnar structures formed due to oblique angle deposition. $\alpha'$ and $\beta'$, are the angles of incoming flux and tilt of columns with respect to substrate normal respectively.

The growth parameters can also lead to the planar stress in the films, which can result in the stress-induced anisotropies. The preferred direction of the magnetization due to this stress – induced anisotropy depends on the nature of stress ($\sigma$) and magnetostriction constant ($\lambda$) of the material, as: $E_{SI} = -K_{SI} \cos^2\theta$, here, $K_{SI}$ is given as $\frac{3}{2}\lambda\sigma$ [1], where $E_{SI}$ is the stress induced anisotropy energy and $K_{SI}$ is the stress induced anisotropy constant. It has been reported that high deposition power lead to tensile stress whereas low power induces compressive stress. On the other hand, compressive stress is resulted from the low Ar pressure and vice-versa. Formation of dense or closely packed columnar structures can also lead to the compressive stress and the nature of stress reverse for the low-density structures [3-4] Our deposition conditions, low power and moderate Ar pressure facilitate the induction of compressive stress in the films. The CFB films are positively magnetstrictive and along with compressive stress (negative), provide an easy direction for magnetization to be out of the film plane driven by energy minimization.

**Quantitative analysis of static magnetic properties for the magnetic samples**

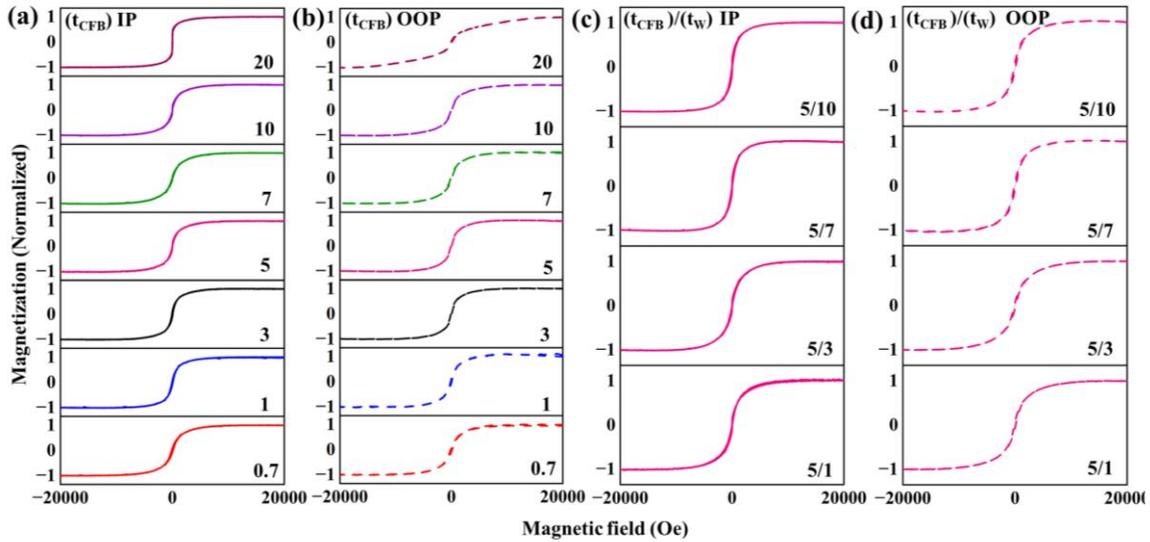

Figure 6S: Normalized IP (solid line) and OOP (dashed line) hysteresis loops for (a-b) Si/CFB ($t_{CFB}$) and (c-d) Si/CFB ($t_{CFB}$ = 5 nm)/W ($t_W$) films respectively, in the field range ±2T. The numbers inside the plots represent the thickness in nm.

Figure 6S shows the in-plane (IP) and out-of-plane (OOP) hysteresis loops for both Si/CFB ($t_{CFB}$) and Si/CFB ($t_{CFB}$ = 5 nm)/W ($t_W$) in the range ±2T.

Figure 7S represents the related magnetic analysis for both Si/CFB ($t_{CFB}$) and Si/CFB ($t_{CFB}$ = 5 nm)/W ($t_W$). We can observe that IP and OOP coercivity ($H_c$) values for Si/CFB ($t_{CFB}$ = 0.7, 1, 3) lie close. This regime is called the interface anisotropy-dominated (IAD) regime in the main manuscript. However, this difference between IP and OOP $H_c$ values increases as we towards higher thicknesses. A somewhat similar trend is observed in the capping layer thickness variation series. Initially, the difference in IP and OOP $H_c$ values is less, but this increases with increasing W capping layer thickness ($t_W$). The IP and OOP saturation field ($H_s$) values for Si/CFB ($t_{CFB}$) follow a similar pattern and have very little to no difference between the two. However, for Si/CFB ($t_{CFB}$ = 20 nm), IP Hs is much lower than OOP Hs, suggesting IP to be the easy direction for the magnetization. These observations imply dominance of IP anisotropy. This regime is termed as shape anisotropy-dominated (SAD) regime. The IP $H_s$ values are slightly lesser than those of OOP for Si/CFB ($t_{CFB}$ = 5 nm)/W ($t_W$). The difference in IP and OOP $H_s$ values is almost similar except for the Si/CFB ($t_{CFB}$ = 5 nm)/W ($t_W$ = 1 nm).

The IP remanence ratio $M_r/M_s$ (%) for the Si/CFB ($t_{CFB}$ = 0.7, 1 and 3 nm) in IAD regime is increasing with thickness. Conversely, these values decrease with thickness in growth induced shape anisotropy dominated (GI-SAD) regime and SAD regime. The OOP $M_r/M_s$ (%) values do not comply with the similar trend. The IP and OOP $M_r/M_s$ (%) values for the Si/CFB ($t_{CFB}$ = 5 nm)/W ($t_W$) films follow a similar trend with a comparable difference between IP and OOP $M_r/M_s$ (%) values.

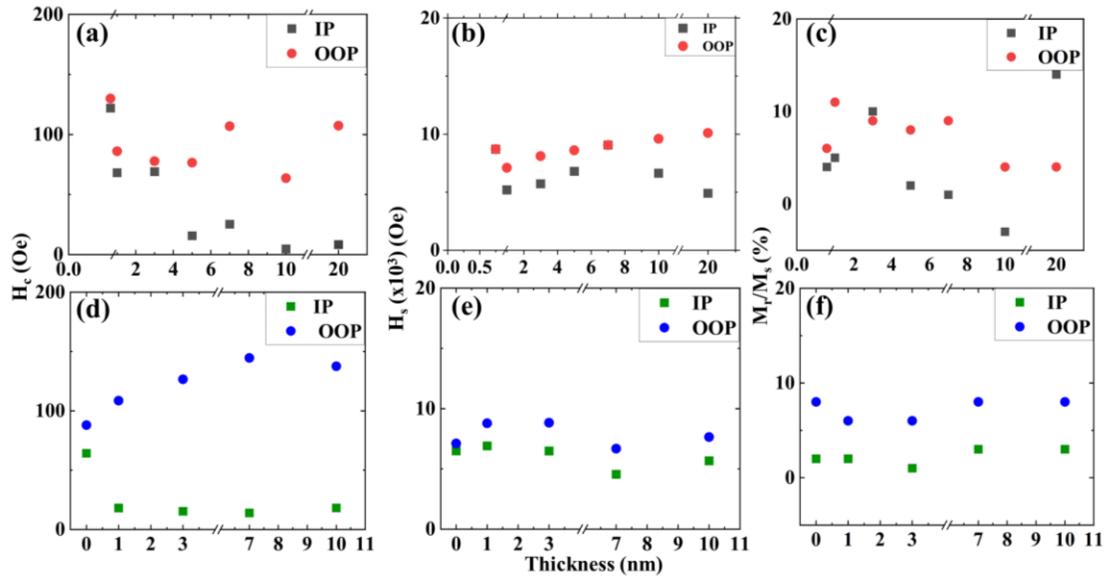

Figure 7S: Variation of coercivity ($H_c$), saturation field ($H_s$), and remanence ratio ($M_r/M_s$ (%)) for (a-c) Si/CFB ($t_{CFB}$) and (e-f) Si/CFB ($t_{CFB}$ = 5 nm)/W ($t_W$), respectively, obtained from the hysteresis loops.

Figure 8S represents the IP inverted loop shapes for the Si/CFB ($t_{CFB}$ = 5 and 10 nm) films. These films exhibit a two-stage reversal. These kinds of loops emerge because of competition between surface uniaxial magnetic anisotropy (UMA) and shape anisotropy of comparable magnitudes. The arrows depict the direction of magnetization with a varying external magnetic field. Points 1, 2, and 3 represent magnetization reversal points at various regions in the hysteresis loop. A representation of the magnetization orientation with respect to the applied field is given in fig. 8S (b). Fig.8S (c) represents the components of magnetization in IP and OOP directions due to GI-SA. We believe that our films have weak IP UMA (shown by double-sided arrows in Fig. 8S(b). In region 1, the component of magnetization in region 1 aligns with the external magnetic field.

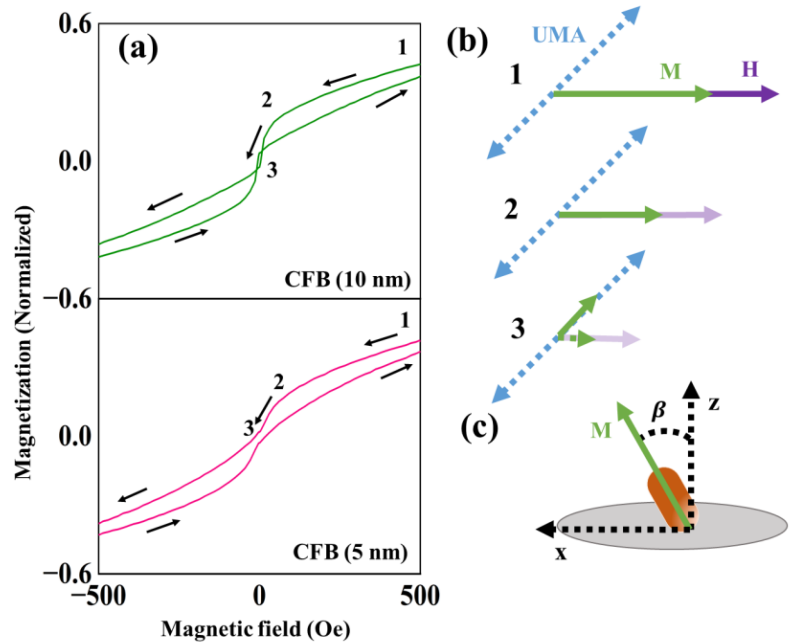

Figure 8S: Conceptual illustration of how magnetization is orienting with respect to applied magnetic field in inverted hysteresis loops. (a) The loops of Si/CFB ($t_{CFB}$ = 5 and 10 nm) are used for this demonstration. (b) Region wise mapping of magnetization orientation at different steps of reversal process with respect to field sweep. (c) IP (x) and OOP (z) component of the magnetization lying along long columnar axis due to GI-SA.

However, moving towards region 2, field magnitude reduces, and it further reduces in region 3, though the in-plane projection of magnetization is substantially small, leading to the sharp jump of magnetization in this region. This kind of behaviour is anomalous for such films.

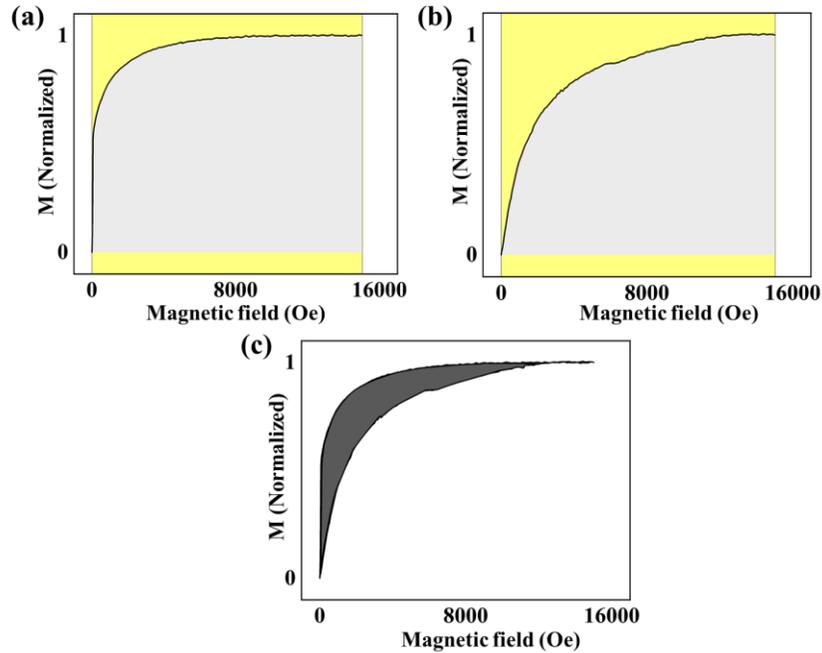

Figure 9S: Visualization of area method. (a-b) The area under the curve for IP and OOP initial curves respective, (c) area (IP) – area (OOP) denoted with the shaded region. The data for CFB ($t_{CFB}$ = 20 nm) sample is used for this illustration.

Figure 9S represents the area method for the calculation of anisotropy values. The area under the IP curve – area under the OOP curve. However, it can be noted that the reverse will also give the true anisotropy values with opposite sign. The selection of this method only depends on the motivation to study a particular phenomenon. The area (OOP curve) – area (IP curve) is chosen in main manuscript to study the PMA energy. However, the calculated anisotropy values from area (IP curve) – area (OOP curve) are also depicted in Fig. 10S.

Figure 10S contains the effective anisotropy $K_{eff}$ values calculated from the aforementioned area method, area (IP curve) – area (OOP curve) with shape anisotropy values for the better impression of the behaviour of the obtained anisotropy values with respect to the inherent shape anisotropy values. This approach is different from the one employed in the main manuscript. This gives us a realization that either of the methods can be used based on the requirement and focus of the study.

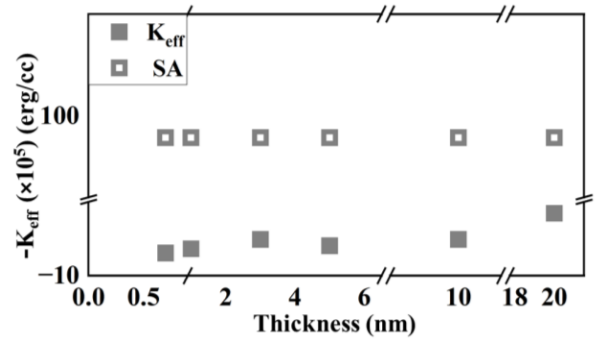

Figure 10S: Effective anisotropy values, $K_{eff}$ for the CFB ($t_{CFB}$) films calculated using area method along with conventional $2\pi M_S^2$ values.

# References

1. C.D.G. B. D. Cullity, Introduction to Magnetic Materials, 2nd ed., Wiley-IEEE Press, 2009.
2. A. Barranco, A. Borras, A.R. Gonzalez-Elipe, A. Palmero, Perspectives on oblique angle deposition of thin films: From fundamentals to devices, Prog Mater Sci 76 (2016) 59–153. https://doi.org/10.1016/j.pmatsci.2015.06.003.
3. S.F. Cheng, P. Lubitz, Y. Zheng, A.S. Edelstein, Effects of spacer layer on growth, stress and magnetic properties of sputtered permalloy film, in: J Magn Magn Mater, 2004: pp. 109–114. https://doi.org/10.1016/j.jmmm.2004.04.027.
4. J.A. Thornton, J. Tabock, D.W. Hoffman, Internal Stresses in Metallic Films Deposited by Cylindrical Magnetron Sputtering*, 1979. https://doi.org/10.1016/0040-6090(79)90550-9.